\documentclass{ifacconf}

\usepackage{graphicx}      
\usepackage{natbib}        
\usepackage{amsmath,amssymb,bm}
\usepackage{tikz}
\usetikzlibrary{arrows.meta,positioning,shapes,decorations.pathreplacing,fit,calc}
\newtheorem{rmk}{Remark}
\begin{document}
\begin{frontmatter}

\title{Multi-Scale Control of Large Agent Populations: From Density Dynamics to Individual Actuation\thanksref{footnoteinfo}}

\thanks[footnoteinfo]{This work has been partially supported by the European Union (EU HORIZON SHARESPACE GA 101092889) and by the MUR PRIN project MENTOR (CUP: E53D23001160006).}

\author[First]{Mario di Bernardo}

\address[First]{Department of Electrical Engineering and Information Technology,
   University of Naples Federico II, and Scuola Superiore Meridionale, Naples, Italy
   (e-mail: mario.dibernardo@unina.it)}

\begin{abstract}
We review a body of recent work by the author and collaborators on
controlling the spatial organisation of large agent populations across
multiple scales. A central theme is the systematic bridging of
microscopic agent-level dynamics and macroscopic density descriptions,
enabling control design at the most natural level of abstraction and
subsequent translation across scales. We make this bridging explicit by
naming the two operations that connect the scales. A
\emph{density-to-agent} (D/A) converter maps a macroscopic control field
to individual agent inputs, while an \emph{agent-to-density} (A/D)
converter reconstructs the macroscopic density from agent measurements.
In this language multi-scale control acquires the structure of a standard
feedback architecture, and different approaches in the literature
correspond to different choices of controller, or of converter. We show
how this multi-scale
perspective provides a unified approach to both \emph{direct control},
where every agent is actuated, and \emph{indirect control}, where few
leaders or herders steer a larger uncontrolled population. The review
covers continuification-based control with robustness under limited
sensing and decentralised implementation via distributed density
estimation; leader--follower density regulation with dual-feedback
stability guarantees and bio-inspired plasticity; optimal-transport
methods for coverage control and macro-to-micro discretisation;
nonreciprocal field theory for collective decision-making; mean-field
control barrier functions for population-level safety; and
hierarchical reinforcement learning for settings where closed-form
solutions are intractable. Together, these results demonstrate the
breadth and versatility of a multi-scale control framework that
integrates analytical methods, learning, and physics-inspired
approaches for large agent populations.
\end{abstract}


\end{frontmatter}

\section{Introduction}

Controlling the collective behaviour of large groups of interacting agents is a central challenge across swarm robotics, traffic management, and synthetic biology \citep{DSouza2023}. When the number of agents~$N$ is large, designing individual control inputs becomes intractable. Mean-field theory overcomes this by reformulating the problem in terms of density functions governed by partial differential equations (PDEs), with complexity independent of~$N$ \citep{Fornasier2014}. The key paradigm is that of \emph{multi-scale control}: closing the feedback loop between macroscopic observables---such as the spatial density of the agents---and microscopic actuation, as illustrated in Fig.~\ref{fig:pipeline}.

Over the past few years, our group has developed a systematic framework based on \emph{continuification}---a three-step pipeline originally proposed in \cite{Nikitin2022} that (i)~lifts the multi-agent system to a macroscopic PDE via a mean-field limit, (ii)~designs a control law at the density level, and (iii)~discretises the result back to agent inputs. We have shown that this pipeline provides a unified approach to two fundamentally different classes of problems:
\begin{itemize}
\item \emph{Direct control}: every agent in a single population receives a control input, and the goal is to shape the collective density towards a desired spatial profile.
\item \emph{Indirect control}: a small number of controllable agents (leaders or herders) must steer a larger population of uncontrolled agents towards a target configuration, through inter-population interactions.
\end{itemize}
In both cases, continuification provides the common multi-scale backbone: the control goal is formulated at the macroscopic density level, but actuation is exerted at the microscopic agent level (see Fig.~\ref{fig:pipeline}). Throughout, we refer to the two operations that connect the scales as \emph{converters} (Remark~\ref{rmk:converters}), a viewpoint that makes the feedback structure of multi-scale control explicit and separates the choice of controller from the choice of conversion. The remainder of this paper reviews the main contributions developed in our group along these two directions.

\begin{figure}[t]
\begin{center}
\resizebox{8.4cm}{!}{%
\begin{tikzpicture}[
  >=Stealth,
  font=\small,
  block/.style={draw,rounded corners=2pt,minimum height=13mm,minimum width=30mm,
                align=center,fill=#1!10,font=\small,text=black},
  eqlabel/.style={font=\scriptsize,text=black!80,align=center},
  arr/.style={->,thick,#1},
]
\node[block=red] (micro) at (0,0) {%
  \textbf{Microscopic}\\[1pt]
  {\scriptsize $\dot{x}_i = \sum_j f(\{x_i,x_j\}_\pi) + u_i$}
};
\node[block=blue] (macro) at (0,3.8) {%
  \textbf{Macroscopic}\\[1pt]
  {\scriptsize $\rho_t + [\rho\,(f*\rho)]_x = q(x,t)$}
};
\node[block=green!60!black] (ctrl) at (6.2,3.8) {%
  \textbf{Control design}\\[1pt]
  {\scriptsize $q = K_p e - [eV^d]_x - [\rho V^e]_x$}
};
\node[block=violet] (disc) at (6.2,0) {%
  \textbf{D/A converter}\\[1pt]
  {\scriptsize $u_i(t) = U(x_i, t)$}
};
\draw[arr=blue!70,line width=1.2pt] (micro.north) -- (macro.south)
  node[midway,left=3pt,eqlabel] {\textbf{Step 1}\\{\scriptsize A/D converter}\\{\scriptsize continuification}};
\draw[arr=green!60!black,line width=1.2pt] (macro.east) -- (ctrl.west)
  node[midway,above=2pt,eqlabel] {\textbf{Step 2}\\{\scriptsize PDE control}};
\draw[arr=violet!70,line width=1.2pt] (ctrl.south) -- (disc.north)
  node[midway,right=3pt,eqlabel] {\textbf{Step 3}\\{\scriptsize discretise}};
\draw[arr=red!70,line width=1.2pt] (disc.west) -- (micro.east)
  node[midway,below=2pt,eqlabel] {{\scriptsize deploy $u_i$}};
\node[draw,dashed,rounded corners=2pt,minimum height=7mm,minimum width=16mm,
      fill=yellow!15,font=\scriptsize,align=center] (ref) at (3.1,5.2) {$\rho^d(x,t)$};
\draw[->,thick,dashed,black!50] (ref.south) -- (ctrl.north);
\end{tikzpicture}
}
\caption{The continuification control pipeline, as introduced in \cite{Maffettone2022CSL,Maffettone2023CDC}. The three steps apply to both direct and indirect control problems. Steps~1 and~3 are the agent-to-density (A/D) and density-to-agent (D/A) \emph{converters} of Remark~\ref{rmk:converters}, which close the loop across scales.}
\label{fig:pipeline}
\end{center}
\end{figure}

\section{The Continuification Pipeline}

The continuification (or continuation) approach, originally proposed in \cite{Nikitin2022}, consists of three steps (Fig.~\ref{fig:pipeline}).

\emph{Step~1 -- Continuification.} A mean-field limit maps the microscopic agent dynamics, e.g.
\begin{equation}\label{eq:micro}
\dot{x}_i = \sum_{j=1}^N f\!\left(\{x_i,x_j\}_\pi\right) + u_i, \quad i=1,\ldots,N,
\end{equation}
where $f$ is a pairwise interaction kernel and $u_i$ is the velocity control input, to a macroscopic mass conservation PDE
\begin{equation}\label{eq:PDE}
\rho_t + [\rho\, (f * \rho + U)]_x = 0,
\end{equation}
where $\rho(x,t)$ is the agents' density, $U(x,t)$ is the macroscopic control velocity field, and `$*$' denotes convolution.

\emph{Step~2 -- Macroscopic control design.} A control input is designed at the PDE level so that $\rho(x,t)$ tends to some desired configuration encoded by the \emph{target} density $\rho^d(x,t)$.

\emph{Step~3 -- Discretisation.} The macroscopic control action is sampled back to the agent level so that each agent receives $u_i(t) = U(x_i,t)$.

\begin{rmk}[Scale converters]\label{rmk:converters}
It is useful to give a name to the two operations that connect the scales in Fig.~\ref{fig:pipeline}. We refer to the map from the macroscopic control field $U(\cdot,t)$ to the individual agent inputs $u_i(t)$ as a \emph{density-to-agent} (D/A) converter, and to the map from a measured agent configuration $\{x_i(t)\}_{i=1}^N$ to a macroscopic density $\hat\rho(\cdot,t)$ as an \emph{agent-to-density} (A/D) converter; the naming deliberately echoes the analogue--digital converters of signal processing. With this terminology, multi-scale control acquires the structure of a standard feedback architecture, comprising a macroscopic controller acting on the density error, a D/A converter on the forward path and an A/D converter on the feedback path. Different approaches then correspond to different choices of controller, or of converter. Spatial collocation (Step~3) and semi-discrete optimal transport are alternative D/A converters, whereas the mean-field limit (Step~1) and kernel density estimation are alternative A/D converters. Note that these last two act at different times. The mean-field limit is an offline modelling device that establishes the existence of a macroscopic description, whereas density estimation is executed online at every sampling instant and is what actually closes the loop across scales. We adopt this terminology throughout.
\end{rmk}

The critical insight is that this pipeline applies regardless of whether $u_i$ acts on every agent (\emph{direct control}) or only on a subset of leaders/herders that influence the remaining population through interaction forces (\emph{indirect control}). In both cases, the control design is performed on a PDE describing the density of the controlled population, and the resulting macroscopic action is discretised to obtain feasible agent-level inputs. What changes between problem classes is the macroscopic model and the controller; the converters of Remark~\ref{rmk:converters} are unchanged.


\section{Direct Control}

In the direct control setting, all $N$ agents are actuated and the objective is to steer the population density $\rho$ towards a desired profile~$\rho^d$. In \cite{Maffettone2022CSL}, we introduced a continuification-based control law for agents interacting on a ring. The macroscopic control source
\begin{equation}\label{eq:q}
q = K_p\, e - [e\,V^d]_x - [\rho\, V^e]_x, \quad e = \rho^d - \rho,
\end{equation}
where $V^d = f*\rho^d$ and $V^e = f*e$, is shown to guarantee global asymptotic convergence to~$\rho^d$ via Lyapunov analysis. The macroscopic control velocity $U$ is recovered from $[\rho\, U]_x = -q$ and discretised to each agent as $u_i = U(x_i,t)$ via spatial collocation, the simplest instance of a D/A converter.

Since $V^e = f*e$ requires global knowledge of the error field, in \cite{Maffettone2023CDC} we studied agents with finite sensing radius~$\Delta$ and proved \emph{semiglobal asymptotic stability}: for any compact set of initial conditions, choosing $K_p$ sufficiently large ensures convergence. Bounded convergence is also established under spatio-temporal velocity disturbances and interaction kernel perturbations, with the residual steady-state error made arbitrarily small by increasing~$K_p$.

A further step towards full scalability was taken in \cite{DiLorenzo2025CSL}, where we replaced centralised density knowledge with a \emph{distributed estimation} scheme, that is, with a decentralised realisation of the A/D converter. Each agent~$i$ maintains a local estimate $\hat\rho^{(i)}$ of the global density, constructed from the agents' own positions via kernel density evaluation and updated through a proportional-integral consensus protocol over a communication graph. The local estimates converge to the true density with bounded error over any strongly connected graph, so that each agent can compute its own control input from its local estimate alone, closing the loop across scales without any centralised computation. The decentralised strategy matches the performance of its centralised counterpart while relying only on local communication.

In \cite{Maffettone2024TCST}, we validated the continuification pipeline experimentally on a physical swarm of mobile robots operating in a mixed-reality environment. The density is estimated online from robot positions and the macroscopic control is discretised in real time, so that both converters are realised in hardware. Experiments confirm the effectiveness of the approach for up to 100 robots tracking multimodal target distributions, demonstrating that continuification control is not merely a theoretical construct but a deployable technology.
The experimental platform merits further comment. In \cite{Maffettone2024TCST}, the mixed-reality setup couples a small number of physical differential-drive robots with a larger virtual population, all interacting in real time through a shared spatial domain. The macroscopic density is estimated online from the joint physical-virtual positions and the continuification control law~\eqref{eq:q} is discretised at each sampling step. In experiments with up to 100~agents tracking unimodal, bimodal, and time-varying target distributions, the closed-loop density converges reliably to~$\rho^d$ with errors consistent with the theoretical predictions of~\cite{Maffettone2023CDC}. This validation confirms that the continuification pipeline is robust to the delays, quantisation errors, and communication imperfections inherent in a physical swarm implementation.

\subsection{Optimal transport for density control}

When the target density~$\rho^d$ varies in time, the continuification control law~\eqref{eq:q} may produce velocity fields that are neither mass-preserving nor energy-efficient. In \cite{Napolitano2026Auto}, we reformulated the direct control problem through \emph{optimal transport} (OT) theory. Given a current density~$\rho(\cdot,t)$ and a desired density~$\rho^d(\cdot,t+\Delta t)$, one solves for the transport map~$T$ minimising
\begin{equation}\label{eq:OT}
\inf_{T:\,T_\#\rho = \rho^d} \int \|x - T(x)\|^2\,\rho(x)\,dx,
\end{equation}
where $T_\#\rho$ denotes the push-forward of~$\rho$ through~$T$. The resulting velocity field $U^{\mathrm{OT}}(x,t) = (T(x)-x)/\Delta t$ is mass-preserving by construction and minimises the kinetic energy of the transport. For one-dimensional populations, closed-form solutions via the quantile function are derived, while for higher dimensions entropy-regularised formulations yield smooth, computationally tractable plans. The OT-based approach provides a principled alternative to the Lyapunov-based law~\eqref{eq:q} when optimality of the transport plan is desired, and naturally interfaces with the continuification pipeline through Step~3 discretisation: each agent receives $u_i = U^{\mathrm{OT}}(x_i,t)$.

In the language of Remark~\ref{rmk:converters}, the semi-discrete OT construction is an \emph{optimality-based D/A converter}, replacing spatial collocation with the solution of~\eqref{eq:OT}. In the semi-discrete case the domain is partitioned into equal-mass Laguerre regions, one per agent, and at the optimum each agent tracks the barycentre of its own region, so that the agent-level control law is exactly a tracking of the first-order optimality condition. It is worth noting that classical coverage control, in the sense of \cite{Cortes2004}, is recovered as the static, fully actuated special case of this converter; here the target density is time-varying and is prescribed by the macroscopic feedback law upstream rather than being chosen a priori.

The OT perspective also opens a geometric viewpoint: the space of density functions, equipped with the Wasserstein-2 metric, becomes a Riemannian manifold on which the controlled density traces a trajectory. This connection suggests a path towards Wasserstein-space Lyapunov analysis and geometric control design that we are actively pursuing.
\section{Indirect Control}

In indirect control, only a small subset of agents is actuated: these \emph{leaders} or \emph{herders} must steer a larger population of uncontrolled agents through inter-population interactions. We have shown that continuification naturally extends to this setting by lifting the coupled multi-population dynamics to a system of PDEs and designing the macroscopic control for the actuated population alone. The converter structure of Remark~\ref{rmk:converters} is preserved, with the D/A converter acting on the actuated population only, while the A/D converter must now reconstruct the densities of both populations.

\subsection{Leader--follower density control}

In \cite{Maffettone2025TAC}, we formulated the leader--follower density control problem. Controllable leaders (density~$\rho_L$) steer uncontrolled followers (density~$\rho_F$) through a repulsive interaction kernel~$f$. The coupled macroscopic model reads
\begin{align}
\partial_t \rho_L + [\rho_L\, u]_x &= 0,\label{eq:L}\\
\partial_t \rho_F + [\rho_F\, (f * \rho_L)]_x &= D\,\rho_{F,xx},\label{eq:F}
\end{align}
where $u(x,t)$ is the leaders' velocity field to be designed and $D>0$ models stochastic follower behaviour. The goal is to find~$u$ such that $\rho_F(x,t) \to \bar\rho_F(x)$.

We derived \emph{feasibility conditions} on the minimum leader mass as a function of the desired follower profile, diffusion, and kernel parameters. A \emph{feedforward} scheme achieves exponential leader convergence with global follower convergence proven via the Poincar\'e--Wirtinger inequality. A \emph{reference-governor} (RG) dual-feedback scheme adapts the leaders' reference as $\hat\rho_L = \bar\rho_L + \alpha(t)\,W$ based on the follower tracking error, with $\alpha(t)\in[0,1]$ ensuring positivity and mass conservation. The RG scheme reduces steady-state errors by up to 90\% compared to feedforward under disturbances. Crucially, the leaders' control~$u$ is ultimately passed through the D/A converter of Remark~\ref{rmk:converters}: each leader agent receives $u_j = u(x_j,t)$.

\subsection{Bio-inspired plasticity and heterogeneous populations}

In \cite{Maffettone2026Auto}, we extended the leader--follower framework with \emph{bio-inspired plasticity}: the coupling adapts online, mimicking biological swarms where interaction strengths evolve with experience. The resulting adaptive architecture provides improved robustness to modelling errors and time-varying environments. In \cite{Maffettone2026TAC}, we further generalised the framework to \emph{heterogeneous} populations with varying dynamics and interaction kernels. The key challenge is that density-level control design assumes a homogeneous population, while real swarms exhibit inter-agent variability in speed, sensing radius, and interaction strength. We modelled this variability as a matched perturbation of the nominal macroscopic dynamics and designed a robust control law that guarantees ultimate boundedness of the density tracking error. The error bound is explicit in the heterogeneity level and can be made arbitrarily small by increasing the control gain, at the cost of higher actuation effort. The robust formulation also applies to scenarios with unknown bounded disturbances---for instance unmodelled environmental forces---extending the framework's applicability to real-world conditions where precise model knowledge is unavailable.

\subsection{Shepherding and nonreciprocal field theory}

Shepherding in swarm robotics represents an extreme form of indirect control: $M$ herders must drive $N_T \gg M$ target agents toward a goal region. In \cite{Lama2024PRR}, we established \emph{scaling laws for herdability}: the minimum number of herders satisfies $M^* \sim N_T^\alpha$ with $\alpha < 1$, confirming that sublinear sparse actuation suffices. A critical density threshold below which herdability degrades was identified and explained through percolation of a herdability graph.

In \cite{Lama2025NC}, we developed a \emph{nonreciprocal field theory} for the shepherding problem. The first required step to apply a continuification based approach to the problem. At the microscopic level, each herder~$i$ selects a target via a soft-max rule that approximates the choice of the furthest target from the goal within its sensing radius~$\xi$:
\begin{equation}\label{eq:selection}
\mathbf{T}_i^* = \frac{\sum_{a \in \mathcal{N}_{i,\xi}} e^{\gamma(|\mathbf{T}_a| - |\mathbf{H}_i|)}\,\mathbf{T}_a}{\sum_{a \in \mathcal{N}_{i,\xi}} e^{\gamma(|\mathbf{T}_a| - |\mathbf{H}_i|)}},
\end{equation}
where $\gamma \geq 0$ controls the selection specificity (from centre-of-mass averaging at $\gamma = 0$ to furthest-target selection as $\gamma \to \infty$). The herder then positions itself behind the selected target via the feedback control input
\begin{equation}\label{eq:herder_ctrl}
\mathbf{u}_i = -\!\left(\mathbf{H}_i - \mathbf{T}_i^* - \delta\,\widehat{\mathbf{T}}_i^*\right),
\end{equation}
where $\delta \geq 0$ controls the goal-directedness of the trajectory planning and $\widehat{\mathbf{T}}_i^* = \mathbf{T}_i^*/|\mathbf{T}_i^*|$. The two decision-making parameters $\gamma$ (target selection) and $\delta$ (trajectory planning) introduce nonreciprocal and three-body couplings that have no counterpart in the targets' dynamics.

By deriving a mean-field limit, we obtained a novel set of coupled field equations for the target density~$\rho^T$ and herder density~$\rho^H$:
\begin{align}
\partial_t \rho^T &= \nabla \cdot \!\left[D^T(\rho^T)\,\nabla\rho^T + \tilde{k}^T \rho^T \nabla\rho^H\right], \label{eq:FPtarget}\\
\partial_t \rho^H &= \nabla \cdot \!\left[D^H(\rho^H)\,\nabla\rho^H - v_1(x)\,\rho^H\rho^T - v_2(x)\,\rho^H\nabla\rho^T\right], \label{eq:FPherder}
\end{align}
where $D^A(\rho^A)$ are renormalised diffusivities arising from noise and short-range repulsion. The coupling functions $v_1(x)$ and $v_2(x)$ encode the decision-making: $v_2$ maintains a constant sign reflecting consistent herder-target attraction, while $v_1$ changes sign depending on the herder position relative to the goal, encoding goal-directed shepherding. The bilinear nonreciprocal term $v_1(x)\,\rho^H\rho^T$ is absent in standard nonreciprocal field theories and arises uniquely from the control-oriented, three-body character of the decision-making rules~(\ref{eq:selection})--(\ref{eq:herder_ctrl}). This analysis reveals that decision-making is itself a fundamental source of nonreciprocity, driving phase transitions between homogeneous and confined target configurations. From the viewpoint of Remark~\ref{rmk:converters}, this field theory is what makes an A/D conversion possible at all in the shepherding setting, by supplying the macroscopic description that the decision-making rules would otherwise preclude.

In \cite{DiLorenzo2025EJC}, we applied the continuification pipeline directly to the shepherding problem: the target population is described by a conservation PDE, a macroscopic herding strategy is designed at the density level, and the resulting control is discretised to the herder agents. This enables regulation of arbitrarily large target populations with few herders. In \cite{Napolitano2025TCST}, we proposed a \emph{hierarchical learning-based} architecture that combines high-level planning with low-level learned policies for robust shepherding of stochastic agents under uncertainty and partial observability. We emphasise that learning is used here to approximate controllers and converters that are not available in closed form, rather than to replace the feedback architecture itself.

\subsection{Population-Level Safety}

In realistic deployment scenarios, multi-agent systems must satisfy safety
constraints---collision avoidance with obstacles, confinement to admissible
regions, and inter-agent spacing guarantees---while pursuing their density
control objective. A natural approach within the multi-scale framework is to
lift the \emph{Control Barrier Function} (CBF) methodology to the macroscopic
level, defining barrier functionals directly on the population density rather
than on individual agent states. The resulting \emph{mean-field CBFs}
(MF-CBFs) impose safety constraints on the macroscopic PDE via a minimally
invasive QP-based safety filter, independent of the number of agents. We are
currently developing this framework for both direct and indirect control,
including shepherding in cluttered environments. In a complementary direction,
in \cite{Punzo2025arXiv} we showed that a driving policy trained via deep RL
in a minimal single-obstacle scenario transfers directly to complex
multi-agent settings, producing collision-free trajectories without
retraining.
\section{Conclusions and Open Challenges}
The body of work reviewed in this paper demonstrates that multi-scale control---the systematic bridging of macroscopic density descriptions and microscopic agent actuation---provides a versatile and rigorous framework for controlling large agent populations. Controlling a large population is not a larger control problem; it is a feedback problem posed at the density level, together with the two scale converters of Remark~\ref{rmk:converters}. The continuification pipeline, optimal transport, nonreciprocal field theory, mean-field control barrier functions, and hierarchical reinforcement learning each address different aspects of the problem, and their combination within this converter architecture yields a control toolbox of considerable breadth.

The leading open challenge is to \emph{certify the converters, and not only the controller}. Convergence and safety are today established at the density level, whereas the errors introduced by the D/A and A/D stages at finite~$N$---discretisation error, density estimation bias and bandwidth selection, and the loss of dynamical information incurred by kernel-based reconstruction---remain largely unquantified. Related open problems include tightening the formal closed-loop guarantees when estimation and control interact across scales, extending the safety framework to richer classes of constraints and system dynamics, developing decision-aware continuum models for populations whose agents make discrete choices, and understanding when analytical, learning-based, and transport-theoretic methods should be combined rather than used in isolation. On the application side, the generality of the multi-scale paradigm invites validation beyond swarm robotics, in domains where large interacting populations and sparse actuation are the norm.

\begin{ack}
The author thanks all the co-authors of the papers presented here for their contributions. During the preparation of this work the author used AI tools to assist with manuscript  formatting and revision.
\end{ack}


\begin{thebibliography}{xx}

\bibitem[{Cort\'es et~al.(2004)}]{Cortes2004}
J.~Cort\'es, S.~Mart\'inez, T.~Karatas, and F.~Bullo.
\newblock Coverage control for mobile sensing networks.
\newblock \emph{IEEE Transactions on Robotics and Automation}, 20(2):243--255, 2004.

\bibitem[{Di~Lorenzo et~al.(2025a)}]{DiLorenzo2025EJC}
B.~Di~Lorenzo, G.C. Maffettone, and M.~di~Bernardo.
\newblock A continuification-based control solution for large-scale shepherding.
\newblock \emph{European Journal of Control}, 86(A):101324, 2025.

\bibitem[{Di~Lorenzo et~al.(2025b)}]{DiLorenzo2025CSL}
B.~Di~Lorenzo, G.C. Maffettone, and M.~di~Bernardo.
\newblock Decentralized continuification control of multi-agent systems via distributed density estimation.
\newblock \emph{IEEE Control Systems Letters}, 9:1580--1585, 2025.

\bibitem[{D'Souza et~al.(2023)}]{DSouza2023}
R.M. D'Souza, M.~di~Bernardo, and Y.-Y. Liu.
\newblock Controlling complex networks with complex nodes.
\newblock \emph{Nature Reviews Physics}, 5:250--262, 2023.

\bibitem[{Fornasier and Solombrino(2014)}]{Fornasier2014}
M.~Fornasier and F.~Solombrino.
\newblock Mean-field optimal control.
\newblock \emph{ESAIM: COCV}, 20(4):1123--1152, 2014.

\bibitem[{Lama and di~Bernardo(2024)}]{Lama2024PRR}
A.~Lama and M.~di~Bernardo.
\newblock Shepherding and herdability in complex multiagent systems.
\newblock \emph{Physical Review Research}, 6(3):L032012, 2024.

\bibitem[{Lama et~al.(2025)}]{Lama2025NC}
A.~Lama, M.~di~Bernardo, and S.H.L. Klapp.
\newblock Nonreciprocal field theory for decision-making in multi-agent control systems.
\newblock \emph{Nature Communications}, 16:8450, 2025.

\bibitem[{Maffettone et~al.(2022)}]{Maffettone2022CSL}
G.C. Maffettone, A.~Boldini, M.~di~Bernardo, and M.~Porfiri.
\newblock Continuification control of large-scale multiagent systems in a ring.
\newblock \emph{IEEE Control Systems Letters}, 7:841--846, 2022.

\bibitem[{Maffettone et~al.(2023)}]{Maffettone2023CDC}
G.C. Maffettone, M.~Porfiri, and M.~di~Bernardo.
\newblock Continuification control of large-scale multiagent systems under limited sensing and structural perturbations.
\newblock \emph{IEEE Control Systems Letters}, 7:2425--2430, 2023.

\bibitem[{Maffettone et~al.(2024)}]{Maffettone2024TCST}
G.C. Maffettone, L.~Liguori, E.~Palermo, M.~di~Bernardo, and M.~Porfiri.
\newblock Mixed reality environment and high-dimensional continuification control for swarm robotics.
\newblock \emph{IEEE Transactions on Control Systems Technology}, 32(6):2484--2491, 2024.

\bibitem[{Maffettone et~al.(2025)}]{Maffettone2025TAC}
G.C. Maffettone, A.~Boldini, M.~Porfiri, and M.~di~Bernardo.
\newblock Leader--follower density control of spatial dynamics in large-scale multi-agent systems.
\newblock \emph{IEEE Transactions on Automatic Control}, 70(10):6783--6798, 2025.

\bibitem[{Maffettone et~al.(2026a)}]{Maffettone2026Auto}
G.C. Maffettone, A.~Boldini, M.~di~Bernardo, and M.~Porfiri.
\newblock Bio-inspired density control of multi-agent swarms via leader--follower plasticity.
\newblock \emph{Automatica}, accepted for publication, 2026.

\bibitem[{Maffettone et~al.(2026b)}]{Maffettone2026TAC}
G.C. Maffettone, D.~Salzano, and M.~di~Bernardo.
\newblock Robust macroscopic density control of heterogeneous multi-agent systems.
\newblock \emph{IEEE Transactions on Automatic Control}, submitted, 2026.

\bibitem[{Napolitano et~al.(2025)}]{Napolitano2025TCST}
I.~Napolitano, S.~Covone, A.~Lama, F.~De~Lellis, and M.~di~Bernardo.
\newblock Hierarchical learning-based control for multi-agent shepherding of stochastic autonomous agents.
\newblock \emph{IEEE Transactions on Control Systems Technology}, submitted, 2025.

\bibitem[{Napolitano and di~Bernardo(2026)}]{Napolitano2026Auto}
I.~Napolitano and M.~di~Bernardo.
\newblock Optimal transport for time-varying multi-agent coverage control.
\newblock \emph{Automatica}, submitted, 2026.

\bibitem[{Nikitin et~al.(2022)}]{Nikitin2022}
D.~Nikitin, C.~Canudas-de-Wit, and P.~Frasca.
\newblock A continuation method for large-scale modeling and control: from {ODE}s to {PDE}, a round trip.
\newblock \emph{IEEE Transactions on Automatic Control}, 67(10):5118--5133, 2022.

\bibitem[{Punzo et~al.(2025)}]{Punzo2025arXiv}
C.~Punzo, I.~Napolitano, C.~Tomaselli, and M.~di~Bernardo.
\newblock Decentralized shepherding of non-cohesive swarms through cluttered environments via deep reinforcement learning.
\newblock \emph{arXiv preprint arXiv:2511.21405}, 2025.

\end{thebibliography}
\end{document}